\documentclass[aps,ams,amsmath,prl,twocolumn,superscriptaddress,longbibliography]{revtex4-1}
\usepackage{graphics}
\usepackage{graphicx}
\usepackage{epsfig}
\usepackage{amsmath}
\usepackage{amssymb}
\usepackage{amsfonts}
\usepackage{bm}
\usepackage{color}
\usepackage{bbm}
\usepackage{hyperref}
\usepackage[normalem]{ulem}
\usepackage{comment}
\usepackage{soul}

\begin{document}

\title{Hydrodynamic magnetoresistance in graphene Corbino devices}

\author{Alex Levchenko}
\affiliation{Department of Physics, University of Wisconsin-Madison, Madison, Wisconsin 53706, USA}

\author{Songci Li}
\affiliation{Department of Physics, University of Wisconsin-Madison, Madison, Wisconsin 53706, USA}

\author{A. V. Andreev}
\affiliation{Department of Physics, University of Washington, Seattle, Washington 98195, USA}

\date{September 8, 2022}

\begin{abstract}
We study hydrodynamic electron magnetotransport  in graphene  devices.
We show that in these systems a distinct mechanism of magnetoresistance appears, which is absent in systems with Galilean-invariant electron liquid. The resulting magnetoresistance depends on the intrinsic conductivity and viscosity of the electron liquid, and becomes especially pronounced near charge neutrality. 
We obtain analytic expressions for magnetoransport coefficients of Corbino devices, and obtain estimates for the electrical and thermal magnetoresistances for monolayer and bilayer systems at charge neutrality.  Magnetoresistance becomes strong (of order $100\,  \%$) at relatively weak fields, at which the kinetic coefficients of the electron liquid are practically unaffected by the magnetic field.
\end{abstract}

\maketitle

\textit{Introduction}. Much of the recent interest in the electronic transport properties in high-mobility two-dimensional electron systems (2DES) concerns the possibility of electron hydrodynamic behavior~\cite{Spivak2010,Narozhny2017,Lucas2018,Levchenko2020,PoliniGeim2020}. This transport regime can be realized in samples of sufficient purity and only in a certain range of temperatures where the mean free path due to electron collisions, $\ell_{ee}$, becomes short compared to other relevant length scales~\cite{Gurzhi1968,AKS2011}. Various experiments in graphene devices, including transport measurements~\cite{Crossno2016,Ghahari2016,Bandurin2016,Kumar2017,Morpurgo2017,Bandurin2018,Berdyugin2019} and local imaging techniques~\cite{Sulpizio2019,Ku2020,Jenkins2020,Brar2021,Ilani2021},  provided numerous pieces of evidence for realization of hydrodynamic electron flow. 

The Corbino disk geometry~\cite{Corbino,Carver} is an attractive alternative to conventional Hall effect measurements of magnetotransport properties of 2DES in rectangular devices. Thermoelectric transport in graphene Corbino devices was studied experimentally~\cite{Fuhrer2010,Geim2010,Zhao2012,Dean2019,Real2020,Hakonen2021}. Hydrodynamic theory was applied to these systems to identify viscous effects in electron transport~\cite{Tomadin2014,Narozhny2015,Holder2019,Shavit2019,Li2022,Gall2022}. 

Because of the potential character of the flow in Corbino geometry the manifestations of viscosity in hydrodynamic transport are rather unusual. In particular, the viscous force density vanishes. Therefore, the Bernoulli law, which generally works only for ideal liquids, applies in this case~\cite{Faber}. Since in the creeping flow regime, which is realized in linear response, the net force density also vanishes, this means that the density of external force driving the flow vanishes as well. In the context of electron hydrodynamics, expulsion of the external force from the flow produces drops of applied voltage and temperature at the sample boundaries~\cite{Shavit2019,Li2022}, which are proportional to the fluid viscosity. It is interesting to note that the energy dissipation associated with these drops occurs in the bulk of the flow. Indeed, the vanishing of the viscous force in Corbino flow does not imply vanishing of the viscous stress tensor, only of its divergence. The viscous stresses arising from the deformation of fluid elements by the flow produce the required energy dissipation.  

Previous theoretical treatments of hydrodynamic magnetotransport assumed Galilean invariance of the electron liquid \cite{Holder2019,Shavit2019,Xie2017}. Therefore their results do not apply to graphene devices near charge neutrality. Motivated in part by experiments \cite{Zhao2012,Dean2019,Real2020,Hakonen2021}, here we develop a theory of hydrodynamic magnetotransport in graphene Corbino devices near charge neutrality. We show that the mechanism of hydrodynamic magnetoresistance (MR) in this case is qualitatively different from that in systems with Galilean-invariant electron liquid, which is caused primarily by the modification of the hydrodynamic flow by the Lorentz force~\cite{Xie2017}. The difference becomes most striking at charge neutrality, where charge current is decoupled from the hydrodynamic flow in the absence of a magnetic field. In contrast,  in a nonzero field the flow remains coupled to the charge current even at charge neutrality. The mechanism of the coupling can be understood as follows. Though the electric current is caused entirely by the intrinsic conductivity, its magnitude is proportional to the electromotive force  (EMF) acting on the electrons. The latter corresponds to the electric field evaluated in the frame moving with the liquid. Therefore, in the presence of a magnetic field the EMF depends on the flow velocity~\cite{LLV8}.  In turn, the Lorentz force caused by the current affects the flow velocity. The magnetoresistance arising from this mechanism depends both the intrinsic conductivity and the viscosity of the electron liquid.

\begin{figure}[t!]
\includegraphics[width=\linewidth]{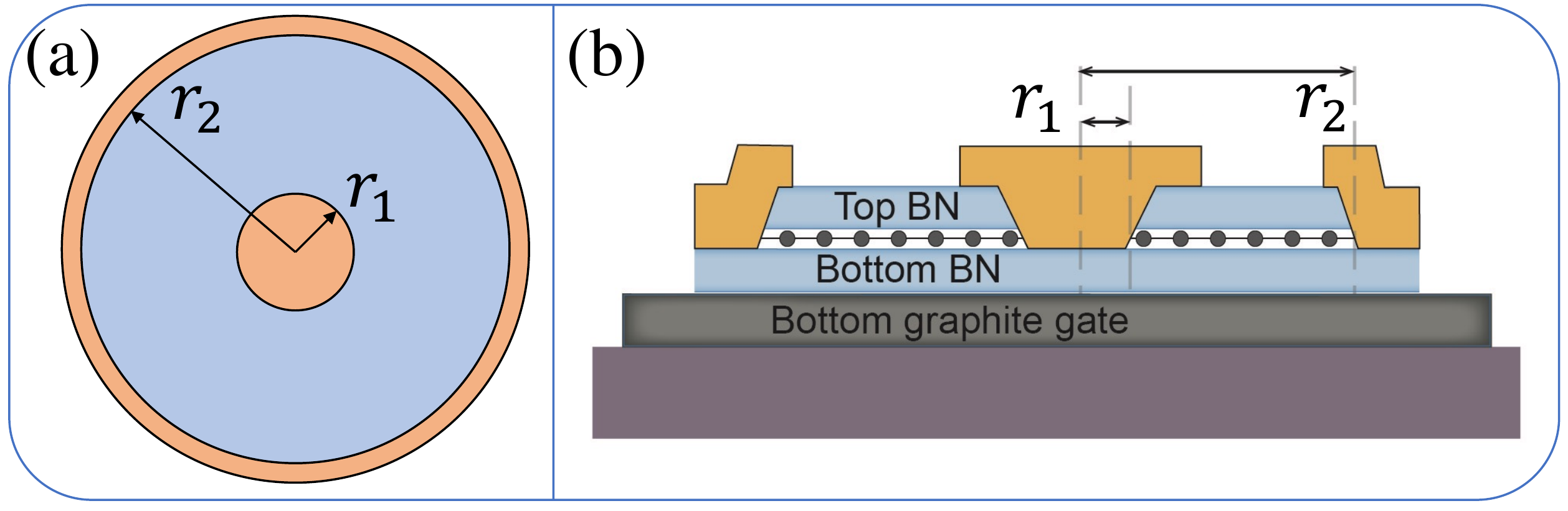}
\caption{Graphene Corbino disk geometry top view is depicted on panel (a). The panel (b) shows the side view of the same device, where a sheet of graphene is encapsulated between the layers of hexagonal boron nitride. The whole device is deposited on the substrate with added gate control.}
\label{Fig-Corbino}
\end{figure}

\textit{Hydrodynamic description}. We consider a Corbino disk geometry with radii $r_1$ and $r_2$, and aspect ratio $p=r_2/r_1>1$, see Fig.~\ref{Fig-Corbino}. The interior of the disk $r\in[r_1,r_2]$ is assumed to be either graphene monolayer (MLG) or bilayer (BLG); more generally, trilayer and multilayers are also possible. The device is subjected to an external magnetic field $\bm{H}=H\hat{\bm{z}}$ applied perpendicular to the $xy$-plane of the disk. We consider a setup, in which a voltage $V$ and a temperature difference $\Delta T $ are applied between the inner and outer electrodes. The electric and heat currents arising in response are denoted by, respectively,  $I=eI_n$ and $I_Q=TI_s$ (for convenience we also introduce particle $I_n$ and entropy $I_s$ currents, where $e$ is the electron charge and $T$ is the temperature). 

The $2\times 2$ thermoelectric resistance matrix $\hat{\mathcal{R}}$ can be determined 
by equating the Joule heat $\mathcal{P}$ to the rate of energy dissipation in the flow  
\begin{equation}
\mathcal{P}=\vec{\mathcal{I}}^{\mathbb{T}}\hat{\mathcal{R}}\vec{\mathcal{I}}, \quad \vec{\mathcal{I}}^{\mathbb{T}}=(I, I_Q). 
\end{equation}
Here we introduced the two-component column vector of currents $\vec{\mathcal{I}}$, with the superscript $\mathbb{T}$ denoting transposition.
The linear response electrical ($R_{\text{el}}$) and thermal ($R_{\text{th}}$) resistances,   
\begin{equation}\label{eq:R-Rth}
R_{\text{el}}(H)=(V/I)_{\Delta T=0}, \quad R_{\text{th}}(H)=(\Delta T/I_Q)_{I=0},
\end{equation}
can be expressed in terms of the matrix elements of $\hat{\mathcal{R}}$ as follows: $R_{\text{el}}(H)=\mathrm{Det}\hat{\mathcal{R}}/\mathcal{R}_{22}$, and $R_{\text{th}}(H)=T\mathcal{R}_{22}$. The off-diagonal elements, $\mathcal{R}_{12}=\mathcal{R}_{21}$, define thermoelectric response.  The  Seebeck coefficient is given by $S(H)=-( V/\Delta T)_{I=0}=-\mathcal{R}_{12}/(T\mathcal{R}_{22})$. From the Onsager relation, $S=\Pi/T$, we can easily determine the Peltier coefficient $\Pi=( I_Q/ I)_{\Delta T=0}$.

The dissipated power in an electron flow, 
\begin{equation}\label{eq:P}
\mathcal{P}=\frac{1}{2}\int\Sigma_{ij}(\partial_iu_j+\partial_ju_i)d^2r+\int \vec{\bm{\mathcal{X}}}^\mathbb{T}\hat{\Upsilon}\vec{\bm{\mathcal{X}}}d^2r,
\end{equation}
comprises of two independent contributions. The first term accounts for the viscous dissipation arising from the hydrodynamic transport mode. In it the stress tensor is given by~\cite{LLV6}
\begin{equation}\label{eq:Sigma}
\Sigma_{ij}=\eta(\partial_iu_j+\partial_ju_i)+(\zeta-\eta)\delta_{ij}\partial_ku_k,
\end{equation}
where $\bm{u}(\bm{r})$ is the hydrodynamic velocity, while $\eta$ and $\zeta$ are, respectively, the shear and bulk viscosities. The form of $\Sigma_{ij}$ in Eq.~\eqref{eq:Sigma} is written for two spatial dimensions. Note also that in Eq.~\eqref{eq:P} the summation over the repeated indices is implicit.  The second term in Eq.~\eqref{eq:P} captures the entropy production rate due to transport in the relative mode, i.e., charge and energy transport relative to the liquid. In Eq.~\eqref{eq:P} the column-vector of thermodynamic forces $\vec{\bm{\mathcal{X}}}$ consists of the electromotive force (EMF) and the temperature gradient. It can be written in the form~\cite{LLV8}
\begin{equation}
\vec{\bm{\mathcal{X}}}=\vec{\bm{X}}-\frac{e}{c}[\bm{u}\times\bm{H}]\vec{\Xi},\quad \vec{\Xi}^\mathbb{T}=(1,0). 
\end{equation} 
The first piece in the expression above is given by a pure gradient, $\vec{\bm{X}}=(-e\bm{E},\bm{\nabla}T)$, where $e\bm{E}$ represents the potential part of the EMF defined by the gradient of electrochemical potential. The second term above describes the contribution to EMF arising from the motion of the liquid in the presence of a magnetic field. The sum of the two contributions corresponds to evaluating the electric field in the frame moving with the liquid~\cite{LLV8}. Finally, the matrix $\hat{\Upsilon}$ in Eq.~\eqref{eq:P} characterizes the dissipative properties of the electron liquid. In the absence of Galilean invariance, it is given by
\begin{equation} \label{eq:Upsilon}
\hat{\Upsilon}=\left(\begin{array}{cc}\sigma/e^2 & \gamma/T \\ \gamma/T & \kappa/ T\end{array}\right)
\end{equation}
and consists of the thermal conductivity $\kappa$, the intrinsic conductivity $\sigma$, and the thermoelectric coefficient $\gamma$, see Refs.~\cite{Mueller2008,Foster2009}. For Galilean-invariant liquids, we have $\sigma =\gamma =0$.
In the consideration below we neglect the dependence of the kinetic coefficients in Eq.~\eqref{eq:Upsilon} on the magnetic field. This approximation assumes that  $\ell_{ee}$ is shorter than the the electron cyclotron radius. We will see that the effects considered below lead to magnetoresistance that becomes strong at very weak magnetic fields, where this approximation is justified.

In order to determine  $\mathcal{P}$ in Eq.~\eqref{eq:P} as a quadratic form of the currents $I$ and $I_Q$ we need to determine the flow pattern, i.e. the spatial profile of $\bm{u}(\bm{r})$ and $\vec{\bm{X}}(\bm{r})$. The hydrodynamic velocity is related to the driving forces by the Navier-Stokes (NS) equation 
\begin{equation}\label{eq:NS}
\eta\nabla^2\bm{u}+\zeta\bm{\nabla}(\bm{\nabla}\cdot\bm{u})=\vec{x}^\mathbb{T}\vec{\bm{X}}+[\bm{j}_n\times\hat{\bm{z}}]/l^2_H,
\end{equation}
which expresses the force balance condition in the bulk of the flow. The first term on the right hand side of Eq. \eqref{eq:NS} describes the potential force on the liquid, which is caused by the temperature and voltage bias. In it we introduced a two-component column vector  $\vec{x}^\mathbb{T}=(n,s)$, whose components are the densities of particles, $n$, and entropy $s$. The last term on the right hand side of Eq.~\eqref{eq:NS} is the Lorentz force, where $ \bm{j}_n$ is the particle current density, and we introduced the magnetic length $l_H=\sqrt{c/|e|H}$. 

The remaining hydrodynamic equations are given by the continuity equations for the particle current $\bm{j}_n$ and entropy current $\bm{j}_s$. Using the column vector notation $\vec{\bm{J}}^\mathbb{T}=(\bm{j}_n,\bm{j}_s)$ they can be written as  
\begin{equation}\label{eq:divJ}
\bm{\nabla}\cdot\vec{\bm{J}}=0,\quad \vec{\bm{J}}=\vec{x}\bm{u}-\hat{\Upsilon}\vec{\bm{\mathcal{X}}}. 
\end{equation}
In the absence of a magnetic field the flow is purely radial. The corresponding thermoelectric matrix, and the distribution of temperature and the electric potential was determined in our previous work~\cite{Li2022}. 
In the present study we extend this analysis to study thermoelectric transport in weak magnetic fields.  

\textit{Magnetoflow pattern}. The magnetohydrodynamic description of electron liquids formulated above applies to any device geometry, and does not assume Galilean invariance of the electron liquid. To make further progress, we specialize to the Corbino geometry and work in polar coordinates  $(r,\phi)$. Owing to the angular symmetry of the Corbino disk, the radial and azimuthal components of the currents and forces do not depend on the polar angle $\phi$. Therefore, NS equation~\eqref{eq:NS} projected onto the radial, $(r)$, and azimuth, $(\phi)$, directions reduces to two coupled equations  
\begin{subequations}
\begin{align}
&(\eta + \zeta )\hat{\Delta}u_r-\vec{x}^\mathbb{T}\vec{X}_r-(j_{n})_\phi/l^2_H=0, \label{eq:NSsa}\\
&\eta\hat{\Delta}u_\phi+(j_{n})_r/l^2_H=0. \label{eq:NSsb}
\end{align}
\end{subequations}
Here $ \hat{\Delta}$ denotes the radial component of the Laplace operator, $\hat{\Delta}=\frac{1}{r}\frac{d}{dr}\left(r\frac{d}{dr}\right)-\frac{1}{r^2}$, and the azimuthal component of the particle current density is given by
\begin{equation}\label{eq:j-phi}
(j_n)_\phi=nu_\phi+\frac{\sigma}{e^2}\frac{u_r}{l^2_H}.
\end{equation}
The continuity equation~\eqref{eq:divJ} for the current densities reads
\begin{equation}\label{eq:continuity}
\vec{J}_r=\frac{\vec{I}}{2\pi r}=\vec{x}u_r-\hat{\Upsilon}\vec{X}_r-\hat{\Upsilon}\vec{\Xi} \, \frac{u_\phi}{l^2_H}, 
\end{equation}
where $\vec{I}^\mathbb{T}=(I_n,I_s)$ is the column vector of particle and entropy currents. In these notations the NS equation for the angular component of the hydrodynamic velocity [Eq. \eqref{eq:NSsb}] reduces to the equation 
\begin{equation}
\hat{\Delta}u_\phi=-\frac{I_n}{2\pi r\eta l^2_H}.
\end{equation}
Its solution is given by 
\begin{equation}\label{eq:uphi}
u_\phi(r)=-\frac{I_nr_1}{4\pi\eta l^2_H}\left(A\rho+\frac{B}{\rho}+\rho\ln\rho \right),
\end{equation}
where we introduced a dimensionless radial coordinate $\rho=r/r_1\in [1,p]$.  The values of the integration constants $A$ and $B$ are determined by the boundary conditions. 
Assuming the standard no-slip boundary condition $u_\phi(r_{1,2})=0$ we obtain  $A=-B=-p^2\ln p/(p^2-1)$.

Next, we analyze the radial part of the NS equation \eqref{eq:NSsa}. For this purpose, we use the continuity equation~\eqref{eq:continuity} to express $\vec{X}_r$ in terms of $u_{r,\phi}$. Then, using Eq. \eqref{eq:j-phi} we obtain
\begin{equation}\label{eq:NS-ur}
( k^2_H - \hat{\Delta})u_r=\frac{\vec{x}^\mathbb{T}\hat{\Upsilon}^{-1}\vec{I}}{2\pi r  (\eta + \zeta)},\quad k^2_H=\frac{1}{l^2}+\frac{\sigma}{e^2(\eta + \zeta) l^4_H}, 
\end{equation}
where the characteristic length scale $l$ is given by 
\begin{equation}\label{eq:l_def}
l^{-2}=\frac{\vec{x}^{\mathbb{T}}\hat{\Upsilon}^{-1}\vec{x}}{(\eta + \zeta)}=\frac{\frac{n^2\kappa}{T}-\frac{2ns\gamma}{T}+\frac{s^2\sigma}{e^2}}
{(\eta + \zeta) \left(\frac{\kappa\sigma}{Te^2}-\frac{\gamma^2}{T^2}\right)}.
\end{equation}
We note that in graphene the bulk viscosity is expected to be negligible~\cite{Principi2016,Narozhny2019}. Therefore, it will be omitted in what follows.

To motivate further approximations it is useful to estimate the order of magnitude of $l$ in different transport regimes. For instance, in the case of MLG in the low density limit close to charge neutrality (Dirac fluid), one gets $l\approx\sqrt{\kappa\eta/Ts^2}\sim l_T$, where $l_T=v/T$ is the thermal de Broglie length. To arrive at this estimation we have used $s,\eta\propto (T/v)^2$ and $\kappa\propto T$ near the neutrality point~\cite{Mueller2009}. In the high density regime (Fermi liquid), one has instead $l\approx\sqrt{\eta/n^2}\sim l_T$, where we have used the estimation of viscosity $\eta \sim n(E_F/T)^2$ in the Fermi liquid regime~\cite{AK1959}. 
The field dependence of $k_H$, and thus $u_r$, is manifested through the parameter $l^2/(\eta l^4_H)\sim (l_T/l_H)^4\ll1$, which is negligible in the hydrodynamic regime since $l_T$ is a microscopic length scale. Consequently, we can set $k_H^2 =l^{-2}$, thereby neglecting the dependence of $u_r$ on the magnetic field. This is accurate within our approximation, in which we neglect the field dependence of the viscosities $\eta$ and $\zeta$, and the kinetic coefficients in matrix $\hat{\Upsilon}$. 

The solution of Eq.~\eqref{eq:NS-ur} consists of the general solution of the homogeneous equation and the particular solution of the inhomogeneous equation. In our approximation, the former  is given by a linear combination of modified Bessel functions of the first and second kinds, $\mathrm{I}_1(r/l)$ and $\mathrm{K}_1(r/l)$. These exponentially decaying and growing solutions of a homogeneous equation are localized on the length $l$ near the inner and outer boundaries and describe deviations of the hydrodynamic flow from that in the bulk. These solutions contribute to the thermoelectric resistance of the contacts. We are interested in the contribution to the resistance matrix due to the hydrodynamic flow in the interior of the disk. The latter corresponds to the particular solution, which is given by 
\begin{equation}\label{eq:ur}
u_r(r)=\frac{1}{2\pi r}\frac{\vec{x}^\mathbb{T}\hat{\Upsilon}^{-1}\vec{I}}{\vec{x}^\mathbb{T}\hat{\Upsilon}^{-1}\vec{x}}. 
\end{equation}
Note that, for given charge and heat currents the radial component of the flow is independent of the magnetic field. In contrast, the azimuthal component of the flow velocity in Eq.~\eqref{eq:uphi} does get modified by the magnetic field. The strength of this modification is characterized by a single dimensionless parameter $\beta=nr^2_1/(2\eta l^4_H)$. This parameter measures the relative strength of the Lorentz and viscous Stokes forces and determines the number of turns the flow makes between the electrodes. For $\beta>1$ the flow swirls around electrodes the integer of $\beta$ times.

\begin{figure}[t!]
\includegraphics[width=0.85\linewidth]{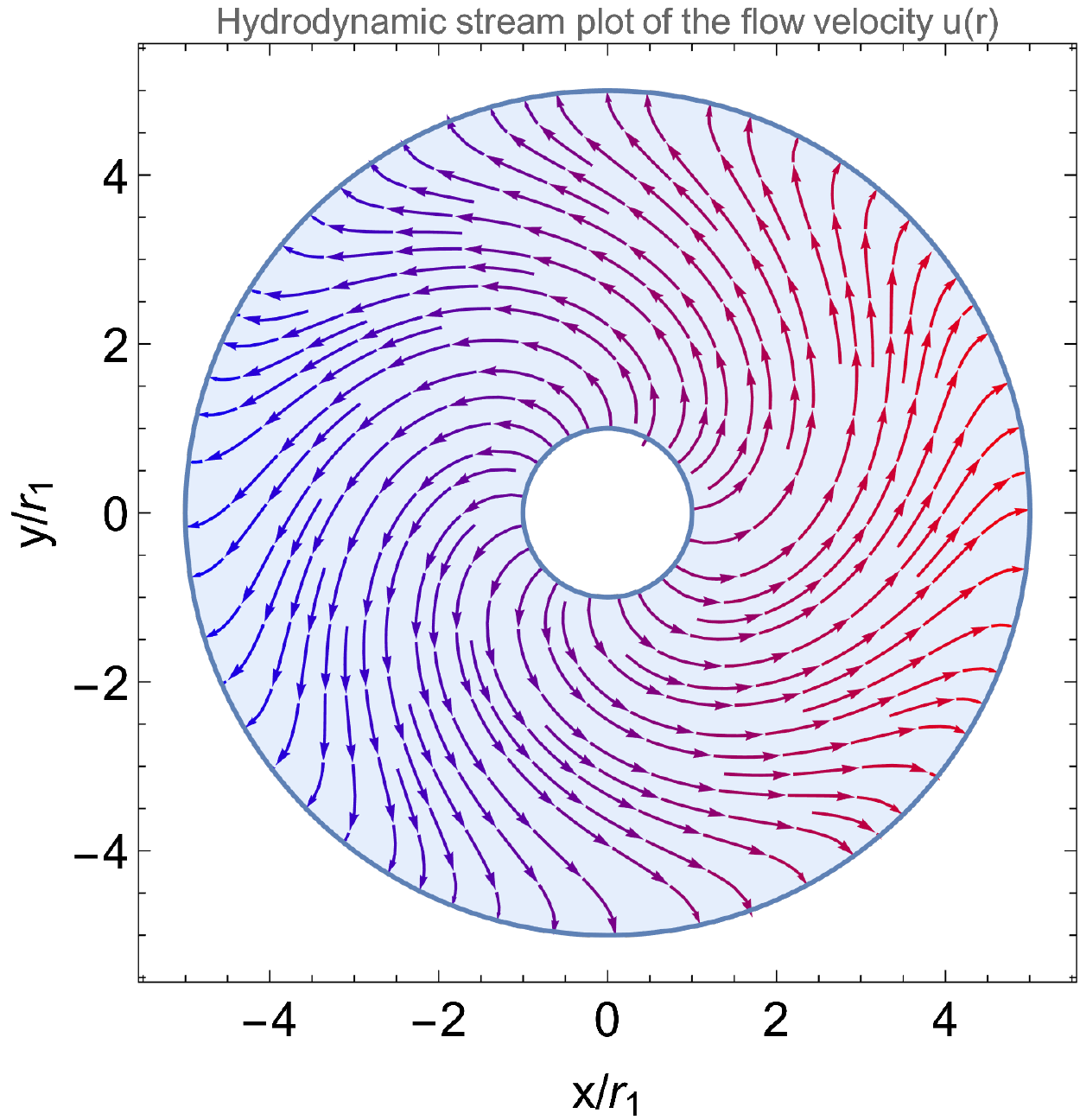}
\caption{
Illustration of hydrodynamic flow of electron liquid in a Corbino device in a perpendicular magnetic field.  The stream plot for the velocity $\bm{u}(\bm{r})$ is generated for the disk with the inner to outer radius ratio $p=r_2/r_1=5$ and non-slip boundary conditions.}
\label{Fig-Flow}
\end{figure}

To illustrate the hydrodynamic magnetoflow in Fig.~\ref{Fig-Flow} we took $\beta=0.5$ and used Eqs.~\eqref{eq:uphi} and~\eqref{eq:ur}  for a particular biasing scenario with $I_s\to0$ at the high density limit $n\gg s$. The velocity in the resulting flow pattern is normalized to  $u_0=I_n/(2\pi r_1n)$. It is worth noting that in the situation corresponding to charge neutrality, $n\to0$, the radial component of the flow velocity in Eq.~\eqref{eq:ur} vanishes, and the charge transport occurs only through the relative mode. In contrast, the azimuthal component $u_\phi$ remains nonzero in the presence of a magnetic field, so that hydrodynamic magnetoflow at charge neutrality is purely vortical.

\textit{Viscous MR}. 
Let us now determine the rate of energy dissipation in this hydrodynamic transport. The derivation naturally breaks down into two steps. First, we use $u_r(r)$ and $u_\phi(r)$ to determine the nonvanishing components of the stress tensor 
\begin{equation}
\Sigma_{rr}=2\eta\frac{\partial u_r}{\partial r},\quad \Sigma_{\phi\phi}=2\eta\frac{u_r}{r},\quad \Sigma_{r\phi}=\eta\left(\frac{\partial u_\phi}{\partial r}-\frac{u_\phi}{r}\right).
\end{equation}
These expressions enable us to calculate the first term in Eq.~\eqref{eq:P}. The second step is to resolve the continuity equation~\eqref{eq:continuity} in order to determine the radial dependence of the forces in the bulk of the flow $\vec{X}_r(r)$. The particular expressions we need read
\begin{subequations}
\begin{align}
&eE_r=\frac{1}{\mathrm{Det}\hat{\Upsilon}}\left[\left(\frac{s\gamma}{T}-\frac{n\kappa}{T}\right)u_r+\frac{\kappa}{T}\frac{I_n}{2\pi r}-\frac{\gamma}{T}\frac{I_s}{2\pi r}\right]+\frac{u_\phi}{l^2_H}, \\ 
&\nabla_rT=\left[\left(\frac{s\sigma}{e^2}-\frac{n\gamma}{T}\right)u_r+\frac{\gamma}{T}\frac{I_n}{2\pi r}-\frac{\sigma}{e^2}\frac{I_s}{2\pi r}\right].
\end{align}
\end{subequations}  
These terms  define the second contribution to $\mathcal{P}$ in Eq. \eqref{eq:P}, which stems from the relative mode. The remaining spatial integrations are elementary, but yield cumbersome expressions. Below, we focus on the regime  near charge neutrality working in leading order in $n/s\ll 1$, and furthermore retain only the leading correction in the magnetic field dependence. 

For the electrical and thermal magnetoresistance we thus find 
\begin{subequations}\label{eq:R-CNP}
\begin{align}
&R_{\text{el}}(H)=R_0\left[1+\frac{\sigma}{e^2}\frac{r^2_2}{\eta l^4_H}f_1(p)\right],\,\, R_0=\frac{\ln p}{2\pi\sigma},\\ 
&R_{\text{th}}(H)=R_{\text{th}}\left[1+\frac{\sigma}{e^2}\frac{r^2_2}{\eta l^4_H}f_2(p)\right],\,\, R_{\text{th}}=\frac{\eta(p^2-1)}{\pi T(r_2s)^2},
\end{align}
\end{subequations}  
where the dimensionless functions of the aspect ratio are 
$f_1(p)=\frac{(p^2-1)^2-4p^2\ln^2p}{8p^2(p^2-1)\ln p},$ and $f_2(p)=\frac{\ln p}{2(p^2-1)}$. 
In order to extract the thermopower we must retain finite density. In the limit $n/s\ll1$ we determine 
\begin{equation}
S(H)=S_0\left[1-\frac{\sigma}{e^2}\frac{r^2_2}{\eta l^4_H}f_2(p)\right],\,\, S_0=\frac{1}{e}\frac{ns}{n^2+\Gamma^2}.
\end{equation}
The characteristic width in the density dependence across the charge neutrality is given by $\Gamma^2=\frac{\sigma}{e^2}\frac{2\eta(p^2-1)}{r^2_2\ln p} $. It  corresponds to electron densities, which are much smaller than the characteristic thermal density $s$. 

\begin{figure}[t!]
\includegraphics[width=\linewidth]{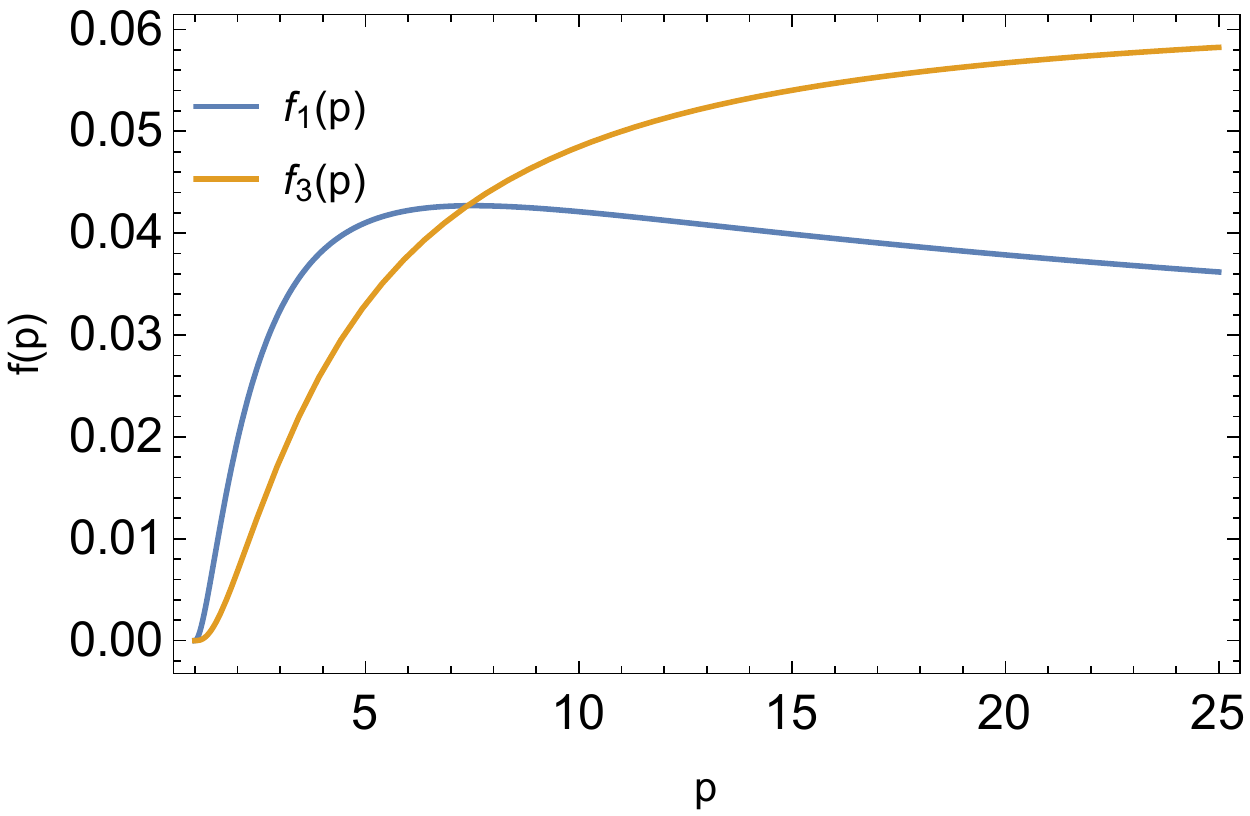}
\caption{Dependence of the dimensionless functions $f_{1,3}(p)$ that define MR in Eqs.~\eqref{eq:R-CNP} and~\eqref{eq:R-FL} on the aspect ratio of the Corbino disk.} 
\label{Fig-f13}
\end{figure}

We see that, just as for Galilean-invariant liquids \cite{Xie2017} MR is positive, but its magnitude is proportional to the intrinsic conductivity and inversely proportional to the viscosity of the electron liquid. The temperature dependence of MR is primarily governed by the fluid viscosity since the intrinsic conductivity is only weakly $T$ dependent. From the weak coupling analysis, it is known that $\sigma$ behaves logarithmically with temperature \cite{Mishchenko2007,Fritz2008,Kashuba2008}. The inverse proportionality of MR to the viscosity is consistent with the earlier results for correlated electron liquids subject to long-range disorder potential~\cite{Xie2017,Patel2017}. This behavior seems to be universal in the hydrodynamic regime. Indeed, extending the above analysis to the opposite limit of high density, $n\gg s$, one finds for MR 
\begin{equation}\label{eq:R-FL}
R_{\text{el}}(H)=R_0+\frac{1}{\pi e^2}\frac{r^2_2}{\eta l^4_H}f_3(p),\,\, R_0=\frac{\eta(p^2-1)}{\pi e^2(r_2n)^2},
\end{equation}
with $f_3(p)=\frac{p^2-1}{16p^2}\left[1-\frac{4p^2\ln^2p}{(p^2-1)^2}\right]$. We note  that the functions $f_1(p)$ and $f_3(p)$, which characterize the magnitude of MR in Eqs.~\eqref{eq:R-CNP} and \eqref{eq:R-FL}, exhibit sensitive dependence on the aspect ratio $p$. As illustrated in Fig.~\ref{Fig-f13} these functions range between zero at $p\to 1$ and practically saturate for $p>10$. 

It is worth noting that at charge neutrality magnetoresistance in Eq.~\eqref{eq:R-CNP} reaches a value of order unity at rather small fields, where $l_H^2 \sim r_1 l_T $ and our approximation of neglecting the dependence of the kinetic coefficients in Eq.~\eqref{eq:Upsilon} and $k_H$ in Eq.~\eqref{eq:NS-ur} on the magnetic field still holds. At large densities MR becomes independent of the intrinsic conductivity. In this regime the absence of Galilean invariance becomes inessential.  

In conclusion, to facilitate possible comparison to experiments, we present estimates for relative MR for both electrical and thermal parts. It is convenient to express them in the form $\delta R_{\text{el},\text{th}}(H)=A_{\text{el},\text{th}}H^2$, where the factors $A_{\text{el},\text{th}}(n,T)$ can be measured independently. Exactly at charge neutrality, we determine that their ratio is $A_{\text{el}}/A_{\text{th}}\propto Ts^2/\eta$. For MLG devices, we thus expect $A_{\text{el}}/A_{\text{th}}\propto T^3$. In BLG, $\eta(T)$ has not been microscopically calculated near charge neutrality, since this is a problem of strong-coupling theory. Nevertheless, we can infer the temperature dependence of $\eta$ from the viscosity to entropy density bound conjecture~\cite{Kovtun2005}. It thus suggests $\eta\sim s(T)\sim m^*T$, where $m^*$ is the effective mass of the band structure. Therefore, for BLG we expect $A_{\text{el}}/A_{\text{th}}\propto T^2$.    

\textit{Note added}. During the completion of this manuscript we became aware of the related work Ref. \cite{Gall-CorbinoMR} that also addresses MR in a graphene Corbino disk at the charge neutrality.   

\textit{Acknowledgments}. We thank V. Gall, I. Gornyi, S. Ilani, P. Kim, B. Narozhny, A. Talanov, and J. Waissman for insightful discussions. This work was financially supported by the National Science Foundation Grants No. DMR-2203411 (A. L.) and DMR-1653661 (S. L.), and MRSEC Grant No. DMR-1719797 (A. V. A.). A.L. is grateful to the Abdus Salam ICTP for hospitality where this work was finalized during the meeting on ``Strongly Correlated Matter: from Quantum Criticality to Flat Bands". 

 \bibliography{biblio}

\end{document}